\documentstyle[twoside,epsf]{article}

\def\PRL #1 #2 #3 {Phys.\ Rev.\ Lett.\ {\bf #1}, #2 (#3)}
\def\PRD #1 #2 #3 {Phys.\ Rev.\ D~{\bf #1}, #2 (#3)}
\def\PLB #1 #2 #3 {Phys.\ Lett.\ B~{\bf #1}, #2 (#3)}
\def\NPB #1 #2 #3 {Nucl.\ Phys.\ {\bf B#1}, #2 (#3)}
\def\ZPC #1 #2 #3 {Z.\ Phys.\ C~{\bf #1}, #2 (#3)}

\newcommand{\gtap}{\;{\raise.3ex\hbox{$>$\kern-.75em\lower1ex\hbox{$\sim$}}}\;}
\newcommand{\ltap}{\;{\raise.3ex\hbox{$<$\kern-.75em\lower1ex\hbox{$\sim$}}}\;}

\catcode`\@=11
\long\def\@makefntext#1{
\protect\noindent \hbox to 3.2pt {\hskip-.9pt  
$^{{\eightrm\@thefnmark}}$\hfil}#1\hfill}		%CAN BE USED 

\def\@makefnmark{\hbox to 0pt{$^{\@thefnmark}$\hss}}	%ORIGINAL 
	
\def\ps@myheadings{\let\@mkboth\@gobbletwo
\def\@oddhead{\hbox{}
\rightmark\hfil\eightrm\thepage}   
\def\@oddfoot{}\def\@evenhead{\eightrm\thepage\hfil
\leftmark\hbox{}}\def\@evenfoot{}
\def\sectionmark##1{}\def\subsectionmark##1{}}

\oddsidemargin=\evensidemargin
\newcounter{sectionc}\newcounter{subsectionc}\newcounter{subsubsectionc}
\renewcommand{\section}[1] {\vspace{12pt}\addtocounter{sectionc}{1} 
\setcounter{subsectionc}{0}\setcounter{subsubsectionc}{0}\noindent 
	{\tenbf\thesectionc. #1}\par\vspace{5pt}}
\renewcommand{\subsection}[1] {\vspace{12pt}\addtocounter{subsectionc}{1} 
	\setcounter{subsubsectionc}{0}\noindent 
	{\bf\thesectionc.\thesubsectionc. {\kern1pt \bfit #1}}\par\vspace{5pt}}
\renewcommand{\subsubsection}[1] {\vspace{12pt}\addtocounter{subsubsectionc}{1}
	\noindent{\tenrm\thesectionc.\thesubsectionc.\thesubsubsectionc.
	{\kern1pt \tenit #1}}\par\vspace{5pt}}
\newcommand{\nonumsection}[1] {\vspace{12pt}\noindent{\tenbf #1}
	\par\vspace{5pt}}

\newcounter{appendixc}
\newcounter{subappendixc}[appendixc]
\newcounter{subsubappendixc}[subappendixc]
\renewcommand{\thesubappendixc}{\Alph{appendixc}.\arabic{subappendixc}}
\renewcommand{\thesubsubappendixc}
	{\Alph{appendixc}.\arabic{subappendixc}.\arabic{subsubappendixc}}

\renewcommand{\appendix}[1] {\vspace{12pt}
        \refstepcounter{appendixc}
        \setcounter{figure}{0}
        \setcounter{table}{0}
        \setcounter{lemma}{0}
        \setcounter{theorem}{0}
        \setcounter{corollary}{0}
        \setcounter{definition}{0}
        \setcounter{equation}{0}
        \renewcommand{\thefigure}{\Alph{appendixc}.\arabic{figure}}
        \renewcommand{\thetable}{\Alph{appendixc}.\arabic{table}}
        \renewcommand{\theappendixc}{\Alph{appendixc}}
        \renewcommand{\thelemma}{\Alph{appendixc}.\arabic{lemma}}
        \renewcommand{\thetheorem}{\Alph{appendixc}.\arabic{theorem}}
        \renewcommand{\thedefinition}{\Alph{appendixc}.\arabic{definition}}
        \renewcommand{\thecorollary}{\Alph{appendixc}.\arabic{corollary}}
        \renewcommand{\theequation}{\Alph{appendixc}.\arabic{equation}}
        \noindent{\tenbf Appendix \theappendixc #1}\par\vspace{5pt}}
\newcommand{\subappendix}[1] {\vspace{12pt}
        \refstepcounter{subappendixc}
        \noindent{\bf Appendix \thesubappendixc. {\kern1pt \bfit #1}}
	\par\vspace{5pt}}
\newcommand{\subsubappendix}[1] {\vspace{12pt}
        \refstepcounter{subsubappendixc}
        \noindent{\rm Appendix \thesubsubappendixc. {\kern1pt \tenit #1}}
	\par\vspace{5pt}}

\newcommand{\textlineskip}{\baselineskip=13pt}
\newcommand{\smalllineskip}{\baselineskip=10pt}

\def\eightcirc{
\begin{picture}(0,0)
\put(4.4,1.8){\circle{6.5}}
\end{picture}}
\def\eightcopyright{\eightcirc\kern2.7pt\hbox{\eightrm c}} 

\newcommand{\copyrightheading}[1]
	{\vspace*{-2.5cm}\smalllineskip{\flushleft
	{\footnotesize International Journal of Modern Physics A, #1}\\
	{\footnotesize $\eightcopyright$\, World Scientific Publishing
	 Company}\\
	 }}

\def\abstracts#1#2#3{{
	\centering{\begin{minipage}{4.5in}\baselineskip=10pt\footnotesize
	\parindent=0pt #1\par 
	\parindent=15pt #2\par
	\parindent=15pt #3
	\end{minipage}}\par}} 

\renewenvironment{thebibliography}[1]
	{\frenchspacing
	 \ninerm\baselineskip=11pt
	 \begin{list}{\arabic{enumi}.}
	{\usecounter{enumi}\setlength{\parsep}{0pt}
	 \setlength{\leftmargin 12.7pt}{\rightmargin 0pt} %FOR 1--9 ITEMS
	 \setlength{\itemsep}{0pt} \settowidth
	{\labelwidth}{#1.}\sloppy}}{\end{list}}

\newcounter{itemlistc}
\newcounter{romanlistc}
\newcounter{alphlistc}
\newcounter{arabiclistc}

\newcommand{\fcaption}[1]{
        \refstepcounter{figure}
        \setbox\@tempboxa = \hbox{\footnotesize Fig.~\thefigure. #1}
        \ifdim \wd\@tempboxa > 5in
           {\begin{center}
        \parbox{5in}{\footnotesize\smalllineskip Fig.~\thefigure. #1}
            \end{center}}
        \else
             {\begin{center}
             {\footnotesize Fig.~\thefigure. #1}
              \end{center}}
        \fi}

\newcommand{\tcaption}[1]{
        \refstepcounter{table}
        \setbox\@tempboxa = \hbox{\footnotesize Table~\thetable. #1}
        \ifdim \wd\@tempboxa > 5in
           {\begin{center}
        \parbox{5in}{\footnotesize\smalllineskip Table~\thetable. #1}
            \end{center}}
        \else
             {\begin{center}
             {\footnotesize Table~\thetable. #1}
              \end{center}}
        \fi}

\def\@citex[#1]#2{\if@filesw\immediate\write\@auxout
	{\string\citation{#2}}\fi
\def\@citea{}\@cite{\@for\@citeb:=#2\do
	{\@citea\def\@citea{,}\@ifundefined
	{b@\@citeb}{{\bf ?}\@warning
	{Citation `\@citeb' on page \thepage \space undefined}}
	{\csname b@\@citeb\endcsname}}}{#1}}

\newif\if@cghi
\def\cite{\@cghitrue\@ifnextchar [{\@tempswatrue
	\@citex}{\@tempswafalse\@citex[]}}
\def\citelow{\@cghifalse\@ifnextchar [{\@tempswatrue
	\@citex}{\@tempswafalse\@citex[]}}
\def\@cite#1#2{{$\null^{#1}$\if@tempswa\typeout
	{IJCGA warning: optional citation argument 
	ignored: `#2'} \fi}}

\def\pmb#1{\setbox0=\hbox{#1}
	\kern-.025em\copy0\kern-\wd0
	\kern.05em\copy0\kern-\wd0
	\kern-.025em\raise.0433em\box0}

\def\fnt#1#2{\footnotetext{\kern-.3em
	{$^{\mbox{\scriptsize #1}}$}{#2}}}

\def\fpage#1{\begingroup
\voffset=.3in
\thispagestyle{empty}\begin{table}[b]\centerline{\footnotesize #1}
	\end{table}\endgroup}

\def\runninghead#1#2{\pagestyle{myheadings}
\markboth{{\protect\footnotesize\it{\quad #1}}\hfill}
{\hfill{\protect\footnotesize\it{#2\quad}}}}
\font\tenrm=cmr10
\font\tenit=cmti10 
\font\tenbf=cmbx10
\font\bfit=cmbxti10 at 10pt
\font\ninerm=cmr9

\font\eightrm=cmr8

\def\qed{\hbox{${\vcenter{\vbox{			%HOLLOW SQUARE
   \hrule height 0.4pt\hbox{\vrule width 0.4pt height 6pt
   \kern5pt\vrule width 0.4pt}\hrule height 0.4pt}}}$}}

\begin{document}

\runninghead{Upper Bound on the Scale of Majorana-Neutrino
Mass Generation}{Upper Bound on the Scale of Majorana-Neutrino
Mass Generation}

\normalsize\textlineskip
\thispagestyle{empty}
\setcounter{page}{1}

\copyrightheading{}			%{Vol. 0, No. 0 (1993) 000--000}

\vspace*{0.88truein}

\fpage{1}
\centerline{\bf UPPER BOUND ON THE SCALE OF}
\vspace*{0.035truein}
\centerline{\bf MAJORANA-NEUTRINO MASS 
	GENERATION\footnote{Talk presented at DPF 2000, Ohio State Univ., 
	Columbus, OH, Aug. 9--12, 2000.}}
\vspace*{0.37truein}
\centerline{\footnotesize J.~M.~NICZYPORUK}
\vspace*{0.015truein}
\centerline{\footnotesize\it Department of Physics, University
of Illinois at Urbana-Champaign}
\baselineskip=10pt
\centerline{\footnotesize\it 1110 West Green Street, 
	Urbana, Illinois 61801-3080, USA} 
%\vspace*{0.225truein}
%\publisher{(received date)}{(revised date)}

\vspace*{0.21truein}
\abstracts{
We derive a model-independent upper bound on the scale of Majorana-neutrino
mass generation.  The upper bound is
$4\pi v^2/\sqrt 3 m_\nu$, where $v \simeq 246$ GeV is the weak
scale and $m_\nu$ is the Majorana neutrino mass.  For neutrino masses
implied by neutrino oscillation experiments, all but one of these
bounds are less than the Planck scale, and they are all within a few orders
of magnitude of the grand-unification scale.}{}{}

\textlineskip			%) USE THIS MEASUREMENT WHEN THERE IS
\vspace*{12pt}			%) NO SECTION HEADING

%\vspace*{1pt}\textlineskip	%) USE THIS MEASUREMENT WHEN THERE IS
%\section{General Appearance}	%) A SECTION HEADING
%\vspace*{-0.5pt}
%\noindent

Considerable evidence
has mounted that one or more of the three neutrino species has a nonzero mass,
based on the observation of neutrino oscillations.\cite{Rob}
Since neutrino masses are necessarily associated with physics beyond the 
Standard Model, one would like to know the energy scale at which this new
physics resides.  In this talk (see Ref.~\ref{ourpaper} for further details),
we present a model-independent upper bound on the scale of
Majorana-neutrino mass generation.    

We assume that the neutrino masses are Majorana, unlike the other known 
fermions, which carry electric charge and are therefore forbidden to have 
Majorana masses.  If there is no 
SU(2)$_L\times$U(1)$_Y$-singlet fermion field in nature, 
then neutrino
masses are necessarily Majorana.  
However, even if such a field exists,
its mass is naturally much greater than the weak scale,\cite{G} 
in which case light neutrinos are Majorana fermions.
In the Standard Model, Majorana neutrino masses are forbidden by
an ``accidental'' global $B-L$ symmetry (baryon number minus lepton
number), but there is no reason to expect that this symmetry
is fundamental.

We begin our analysis with the Standard Model, but with a Majorana neutrino 
mass of unspecified origin.  Since the neutrino mass is put in artificially, 
this is only an effective field theory, valid up to some energy scale 
at which it is subsumed by
a deeper theory, which we regard as the scale of Majorana-neutrino mass
generation.  
A simple way to derive an upper bound on the scale at which 
the effective theory breaks down is to impose
unitarity on the tree-level $2\to 2$
scattering amplitudes that grow with energy.\cite{AC}  
In the high-energy limit, $s \gg M_W^2, M_Z^2,
m_\nu^2, m_\ell^2$, the relevant 
zeroth-partial-wave amplitudes (see Fig.~1) are given by
\begin{eqnarray}
a_0\left(\frac{1}{\sqrt 2}\nu_{i\pm} \nu_{j\pm} \to W^+_LW^-_L\right) & 
\sim & \mp\frac{m_{\nu_i} \sqrt s}{8\pi\sqrt 2v^2}\delta_{ij} \label{WW}\\ 
a_0\left(\frac{1}{\sqrt 2}\nu_{i\pm} \nu_{j\pm} 
\to \frac{1}{\sqrt 2}Z^0_LZ^0_L\right) & 
\sim & \mp\frac{m_{\nu_i} \sqrt s}{8\pi v^2}\delta_{ij} \label{ZZ}
\end{eqnarray}
where $v = (\sqrt 2 G_F)^{-1/2} \simeq 246$ GeV
is the weak scale, the indices $i,j$ denote the 
three neutrino mass eigenstates, 
the subscripts on the neutrinos and charged leptons indicate 
helicity $\pm 1/2$, and the subscript on the partial-wave 
amplitude indicates $J=0$.

\begin{figure}[!t]
\begin{center}
\vspace*{0cm}
\hspace*{0cm}
\epsfclipon
\epsfxsize=8.6cm \epsfbox{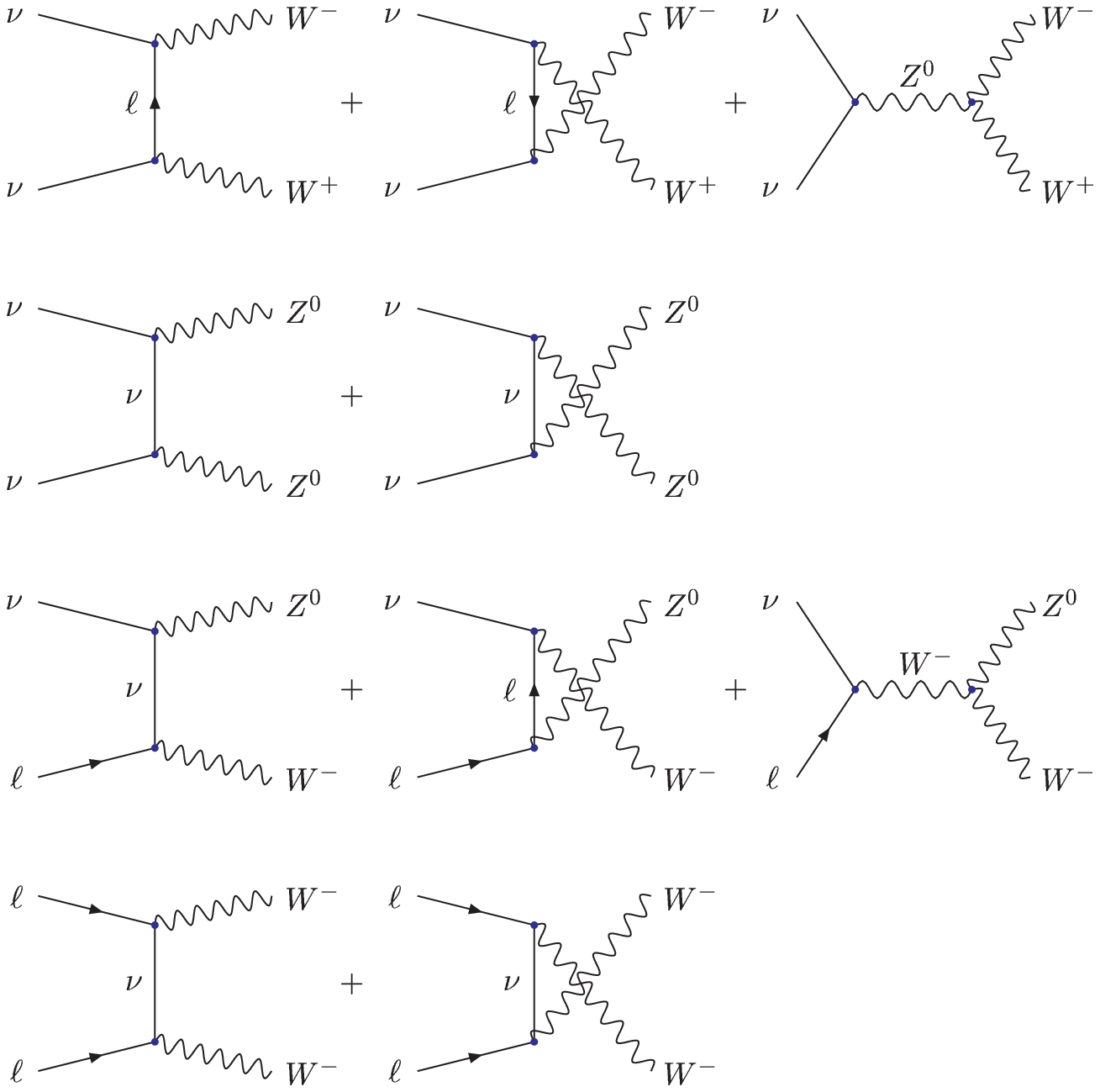}
\vspace*{0cm}
\caption{Feynman diagrams that contribute to the amplitudes in 
Eqs.~(\ref{WW})--(\ref{ZZ}).  The source of the Majorana neutrino mass
is unspecified, so there are no diagrams involving the coupling of the
Majorana neutrino to the Higgs boson.  Unitary gauge is used throughout.}
\label{fig:fig1}
\end{center}
\end{figure}

The strongest bound on the scale of Majorana-neutrino mass generation is
obtained by considering a scattering process which is a linear combination
of the above amplitudes:
\begin{equation}
a_0\left(\frac{1}{2}(\nu_{i+}\nu_{i+} - \nu_{i-}\nu_{i-}) \to 
\frac{1}{\sqrt 3}(W^+_LW^-_L + Z^0_LZ^0_L)\right)
\sim -\frac{\sqrt 3m_{\nu_i} \sqrt s}{8\pi v^2}\;.
\end{equation}
The unitarity condition on inelastic $2\to 2$ scattering amplitudes,
$|a_J| \leq 1/2$,\cite{MVW} implies that the scale of Majorana-neutrino
mass generation is less than the scale
\begin{equation}
\Lambda_{Maj} \equiv \frac{4\pi v^2}{\sqrt 3 m_{\nu}}\;,
\label{BOUND}
\end{equation}
which is inversely proportional to the neutrino mass.  This is our principal
result.

We have analyzed\cite{ourpaper} two explicit models of 
light Majorana-neutrino masses that exemplify
this bound: one with a heavy right-handed neutrino $\nu_R$
(see-saw mechanism), and one with a heavy SU(2)$_L$-triplet Higgs field.  
Unitarity is restored by the heavy
Majorana neutrino $N \approx \nu_R$ in the see-saw model
and by the heavy Higgs scalars in the triplet Higgs model,
whose masses respect the bound in Eq.~(\ref{BOUND}).
In both cases, the bound is saturated when dimensionless 
parameters in these models (Yukawa couplings and/or mass ratios) 
attain their maximum values.  
Indeed, these models can naturally saturate our
bound, Eq.~(\ref{BOUND}), precisely because they 
generate a Majorana-neutrino
mass term of dimension five in the low-energy theory.
In general, saturating unitarity-based bounds is nontrivial;
for example, there is no known model that saturates the 
upper bound on the scale of Dirac-fermion mass generation.\cite{AC,Go}

\begin{table}[t]
%\squeezetable
\caption{Neutrino mass-squared differences from a variety of neutrino
oscillation experiments, and their interpretations.  The last column gives
the upper bound on the scale of Majorana-neutrino mass generation,
Eq.~(\ref{BOUND}), for each interpretation.  Table adapted from Ref.~\ref{Rob}.}
\smallskip
\hbox to\hsize{\hss\vbox{\hbox{\begin{tabular}[4]{lllc}
\hline\noalign{\vskip2pt}\hline\noalign{\vskip2pt}
Experiment&Favored Channel&$\Delta m^2$ ($\rm eV^2$)
&$\Lambda_{Maj}$ $({\rm GeV})<$\\
\noalign{\vskip2pt}\hline\noalign{\vskip2pt}
LSND &$\bar\nu_\mu\to\bar\nu_e$&$0.2-2.0$&$9.8\times10^{14}$\\
Atmospheric&$\nu_\mu\to\nu_\tau$&$3.5\times10^{-3}$&$7.4\times10^{15}$\\
%&$\nu_\mu\to\nu_s$&\multicolumn{2}{l}{Disfavored at $\sim 2\sigma$}\\
 Solar\\
\quad MSW (large angle)
&$\nu_e\to\nu_\mu$ or $\nu_\tau$&
$(1.3-18)\times10^{-5}$&$1.2\times10^{17}$\\
\quad MSW (small angle)&$\nu_e\to$ anything &$(0.4-1)\times10^{-5}$
&$2.2\times 10^{17}$\\
\quad Vacuum&$\nu_e\to\nu_\mu$ or
$\nu_\tau$ &$(0.05-5)\times10^{-10}$&$2.0\times10^{20}$\\
\noalign{\vskip2pt}
\hline
\noalign{\vskip2pt}
\hline
\end{tabular}}}\hss}
\label{massdiffs}
\end{table}

Neutrino oscillation experiments do not measure the neutrino mass, but 
rather the absolute value of the mass-squared difference of two species 
of neutrinos,
$\Delta m^2$.  This implies a lower bound of $m_\nu \ge \sqrt{\Delta m^2}$
on the mass of one of the two participating neutrino species.  
Using Eq.~(\ref{BOUND}), one finds the upper bounds on the scale 
$\Lambda_{Maj}$
given in Table~1 for a variety of neutrino oscillation experiments.
These upper bounds are all within a few orders of magnitude of the Planck 
scale, $G_N^{-1/2} \simeq 1.2\times 10^{19}$ GeV, which is the 
scale before which quantum gravity must become relevant.  However, only the 
vacuum-oscillation interpretation of the solar neutrino deficit yields a 
scale that could be as large as the Planck scale.   In all other cases, 
we find that the physics of Majorana-neutrino mass generation must be below
the Planck scale.  Thus, if these neutrino masses arise from quantum gravity,
then the scale of quantum gravity must be somewhat less than the Planck scale.

The upper bounds on $\Lambda_{Maj}$ are also within a few orders of
magnitude of the grand-unification scale, ${\cal O}(10^{16})$ GeV. 
The LSND and
atmospheric neutrino experiments yield an upper bound on $\Lambda_{Maj}$
slightly below the grand-unification scale, but the scale of Majorana-neutrino
mass generation could be less than the unification scale 
in a grand-unified model.  
For example, in a grand-unified model that makes use of the see-saw 
mechanism, the mass of the heavy Majorana neutrino $N$ could be equal to
a small Yukawa coupling times the vacuum-expectation value of the Higgs 
field that breaks the grand-unified group.

\nonumsection{References}
%\noindent

\end{document}